\documentclass[a4paper,fleqn,usenatbib]{mnras}

\usepackage{newtxtext,newtxmath}

\usepackage[T1]{fontenc}
\usepackage{ae,aecompl}

\usepackage{graphicx}	
\usepackage{amsmath}	
\usepackage{amssymb}	
\usepackage{caption}
\title[Internal gas in 47 Tuc]{Internal gas models and central black hole in 47 Tucanae using millisecond pulsars}

\author[F. Abbate et al.]{
F. Abbate$^{1,2}$\thanks{E-mail: f.abbate@campus.unimib.it},
A. Possenti$^{2}$,
A. Ridolfi$^{3}$, 
P. C. C. Freire$^{3}$,
F. Camilo$^{4}$, \newauthor
R. N. Manchester$^{5}$, and
N. D'Amico$^{2,6}$
\\
$^{1}$Dipartimento di Fisica `G. Occhialini', Universit\`a degli Studi Milano Bicocca, Piazza della Scienza 3, Milano, Italy\\
$^{2}$INAF - Osservatorio Astronomico di Cagliari, Via della Scienza, I-09047 Selargius (CA), Italy\\
$^{3}$Max-Planck-Institut f\"ur Radioastronomie, Auf dem H\"ugel 69, D-53121 Bonn, Germany\\
$^{4}$South African Radio Astronomy Observatory (SARAO), Pinelands, 7405, South Africa\\
$^{5}$CSIRO Astronomy and Space Science, Autralia Telescope National Facility, Box 76, Epping, NSW 1710, Australia \\
$^{6}$Dipartimento di Fisica, Universit\`a degli Studi di Cagliari, SP Monserrato-Sestu km 0,7, 90042 Monserrato (CA), Italy
}

\date{Accepted XXX. Received YYY; in original form ZZZ}

\pubyear{2018}

\begin{document}
\label{firstpage}
\pagerange{\pageref{firstpage}--\pageref{lastpage}}
\maketitle

\begin{abstract}

Despite considerations of mass loss from stellar evolution suggesting otherwise, the content of gas in globular clusters seems poor and hence its measurement very elusive.  One way of constraining the presence of ionized gas in a globular cluster is through its dispersive effects on the radiation of the millisecond pulsars included in the cluster. This effect led Freire et al. in 2001 to the first detection of any kind of gas in a globular cluster in the case of 47 Tucanae.  By exploiting the results of 12 additional years of timing, as well as the observation of new millisecond pulsars in 47 Tucanae, we revisited this measurement: we first used the entire set of available timing parameters in order to measure the dynamical properties of the cluster and the three-dimensional position of the pulsars. Then we applied and tested various gas distribution models: assuming a constant gas density, we confirmed the detection of ionized gas with a number density  of $n= 0.23\pm 0.05$ cm$^{-3}$, larger than the previous determination (at $2\sigma$ uncertainty). Models predicting a decreasing density or following the stellar distribution density are highly disfavoured. We are also able to investigate the presence of an intermediate mass black hole in the centre of the cluster, showing that is not required by the available data, with an upper limit for the mass at $\sim 4000$ M$_{\odot}$.
 
\end{abstract}

\begin{keywords}
globular clusters: individual: 47 Tucanae -- pulsars -- stars: black holes -- stars: kinematics and dynamics -- ISM: kinematics and dynamics
\end{keywords}
\section{Introduction}

Globular clusters (GCs) are known to harbour a very large population of pulsars. Currently 150 pulsars have been found in 28 globular clusters\footnote{For an up-to-date number visit: \href{http://www.naic.edu/\textasciitilde pfreire/GCpsr.html}{http://www.naic.edu/\textasciitilde pfreire/GCpsr.html}}. 
Almost all of them are millisecond pulsars (MSPs) which have been recycled during accretion from a binary companion. 

In globular clusters, the number of low-mass X-ray binaries (and their products, MSPs) per unit stellar mass is much greater than in the Galactic field \citep{1975ApJ...199L.143C}. This results from the very high central densities of the clusters which increases the chance of close stellar encounters. These encounters can lead to the formation of new binaries containing a neutron star where accretion can occur and the neutron star can be recycled. The same processes that form these binaries can also destroy \citep{VerbuntFreire2014} them or the neutron star can ablate the companion with its strong wind. For these reasons many MSPs in globular clusters are isolated.

Thanks to their abundance and to their rotational stability, MSPs can be used as unparalleled probes of the gravitational potential and environment of globular clusters. MSPs have been used to constrain the properties of the parent clusters \citep{Phinney1993, Anderson1993, Freire2003, Prager2017} and to study the presence of intermediate mass black holes (IMBHs) \citep{Kiziltan2017a, Kiziltan2017b, Freire2017, Perera2017, Prager2017}.

47 Tucanae (also known as NGC 104, hereafter 47 Tuc) is one of the most prominent globular clusters. The main properties of the cluster are listed in Table \ref{tab47tuc}. 
The central density has been measured from the proper motion central velocity dispersion $\sigma_{\mu,0}= 0.574 \pm 0.005$ mas yr$^{-1}$ \citep{Watkins2015} and the angular core radius $\theta_c=26.5\pm 1.16$ arcsec \citep{Bellini2017} using equation 1-34 in \cite{Spitzer1987}: 

\begin{equation} \label{vel_disp_eq}
\rho_0=\frac{9\sigma_{\mu,0}^2}{4 \pi \rm{G} \theta_c^2} = (7.5 \pm 0.2) \times 10^{4}\, \rm{M}_{\odot}\, \rm{pc}^{-3};
\end{equation}
which is accurate to $\sim 0.5$ per cent for clusters in which the tidal radius is much larger than the core radius \citep{Freire2017}.
This value is significantly lower than that derived by \cite{Freire2017} because of the significant difference in $\theta_c$. 
We chose to use the value of $\theta_c=26.5\pm 1.16$ arcsec \citep{Bellini2017} as it is the most recent value obtained combining the surface brightness profile with the kinematic information both along the line of sight and in proper motion taken from state-of-the-art Hubble Space Telescope observations.
Radio observations of 47 Tuc led to the discovery of 25 pulsars \citep{Manchester1990, Robinson1995, Camilo2000, Pan2016} and to the phase-coherent timing solution of 23 of them \citep{Freire2001a, Freire2003, Ridolfi2016, Freire2017,Freire2018}. All of these pulsars are MSPs and have spin periods shorter than 8 ms. Fifteen of them are in binary systems. Except for 47 Tuc X, they all reside within 1 arcminute of the centre. \cite{Ridolfi2016} and \cite{Freire2017} also provided the values of the period derivative, the second period derivate and proper motion for 22 pulsars. For 10 of the binary pulsars it was also possible to measure the orbital period derivative.

\begin{table}
\caption{Main properties of the globular cluster 47 Tuc.
 $^1$\protect\cite{McLaughlin2006}, $^2$\protect\cite{Bogdanov2016}, $^3$\protect\cite{Harris2010}, $^4$\protect\cite{Bellini2017}, $^5$\protect\cite{Gratton1997}, $^6$\protect\cite{Gnedin2002},  $^7$\protect\cite{Watkins2015}, $^{8}$\protect\cite{Freire2001b} }

\begin{center}
\begin{tabular}[t]{c c c}
\hline
{Parameter} & {Value} & {References}  \\
\hline
\hline
 Centre RA (J2000)& $00^{\rm h} 24^{\rm m} 05^{\rm s} .67 \pm 0^{\rm s} .07$    &   $^1$ \\
Centre Dec (J2000) &  $-72^{\circ} 04'52".62 \pm 0".26 $  &   $^1$   \\
Distance from Sun &  $4.53 \pm 0.04$ kpc  &   $^2$      \\
Metallicity & $-0.72$ dex & $^3$ \\
Mass &   $(8.4 \pm 0.4) \times 10^5 \rm{M}_{\odot}$ &   $^4$    \\
Tidal radius &   42'.3 (55.4 pc) &     $^3$\\
Core radius &  26".5 (0.58 pc)  &    $^4$ \\
Age &  $10.0 \pm 0.4 $ Gyr&  $^5$     \\
Escape velocity at core &  68.8 km/s  &    $^6$    \\
Central velocity dispersion& $0.574\pm 0.005  $ mas yr$^{-1}$ &  $^7$  \\
Central density&  $75000 \pm 2000\, \rm{M}_{\odot}\, \rm{pc}^{-3}$  & $^{4,7}$       \\
Central ICM density &   $0.067\pm 0.015 \, \rm{cm}^{-3}$ &  $^{8}$     \\
\hline

\end{tabular}\\
\label{tab47tuc}

\end{center}

\end{table}

47 Tuc was the first GC where evidence of the presence of ionized gas was found \citep{Freire2001b}. This discovery was made possible by the study of the dispersion measure (DM) differences between each pulsar. The DM causes a frequency dependent delay of the time of arrival of the pulses and is caused by the presence of free electrons along the line of sight. 

The DM was seen to be higher for pulsars farther along the line of sight compared to ones closer to the observer. This was interpreted as being due to the presence of an ionized component of constant density of the intracluster medium (ICM). Because of the large errors it was not possible to discriminate between various distribution models of the gas.

Despite this detection, there is very little additional evidence of any kind of interstellar medium inside globular clusters. This is a long standing problem in the astrophysics of GCs \citep{Smith1990, VanLoon2006, Barmby2009}. The only certain detection of neutral gas in a globular cluster was made in M15: an HI cloud of $\sim 0.3\, \rm{M}_{\odot}$ and $9 \pm 2 \times 10^{-4}\, \rm{M}_{\odot}$ of dust \citep{Evans2003, Boyer2006, VanLoon2006}. This amount of gas and dust is very small if compared to what is expected to be emitted by the evolved stars of the cluster, i.e. $\sim 10^{-6}\, \rm{M}_{\odot} \rm{yr}^{-1}$ \citep{McDonald2011}. A fast clearing mechanism for the dust is necessary to explain the discrepancy between the observations and the predictions. This clearing mechanism could be caused by pulsar winds \citep{Spergel1991}, fast winds from red giants \citep{Smith2004}, classical novae \citep{Moore2011} or by white dwarfs \citep{McDonald2015a}.

A more detailed modelling of the gas density could in principle be used as a tracer of the origin and evolution of the gas itself. Furthermore, it has been suggested that the distribution of gas could be influenced by the presence of an intermediate mass black hole \citep{Pepe2016}, thus allowing us to put additional constraints on its presence.

Stringent upper limits have been put in the past on the mass of the central IMBH in 47 Tuc between 1000 - 5000 M$_{\odot}$ from both kinematic methods and radio continuum observations \citep{McLaughlin2006, Maccarone2008, Maccarone2010, Lu2011}. Recently a claim of an IMBH of 2200 M$_{\odot}$ was put forward \citep{Kiziltan2017a, Kiziltan2017b} using pulsar observations. However, using updated results, \cite{Freire2017} deemed the claim unnecessary; the same result was obtained by \cite{2018arXiv180703307M} using detailed measurements of the normal stars in the cluster; and a similar conclusion was derived for a larger number of globular clusters from radio continuum surveys \citep{2018ApJ...862...16T}. An independent method for testing for the presence of the IMBH might help to solve this question.

The aim of this paper is to test various distribution models for the ionized gas inside the globular cluster 47 Tuc using the new timing results presented in \cite{Ridolfi2016}, \cite{Freire2017} and \cite{Freire2018}: they were obtained from a much longer data-span ($16$ yrs as compared to $4$ yrs) than that available at the time of the original detection \citep{Freire2001b}. 
The analysis is made using a Markov Chain Monte Carlo (MCMC) algorithm first used to determine the dynamical parameters of Terzan 5 \citep{Prager2017}. Since the core radius and the velocity dispersion of 47 Tuc are well constrained thanks to optical observations, this algorithm can be used to accurately measure the line-of-sight position of the pulsars and to test the presence of an IMBH using the equations described in Section \ref{section_theory}. The algorithm itself is described in Section \ref{section_mcmc}. With the three-dimensional positions of the pulsars and their measured values of DM we test the presence of ionized gas with different distributions in Section \ref{section_gas}. In Section \ref{section_discussion} and Section \ref{section_conclusion} we discuss the results and derive the conclusions.

\section{Theory}\label{section_theory}

GCs are typically modeled according to the King potential \citep{King1962}.  As shown in \cite{Miocchi2013} this model provides an excellent fit for the surface brightness profile of 47 Tuc. Using this model we can predict the values of the velocity dispersion, accelerations and jerks of the pulsars and compare them with the observed values in order to derive the line-of-sight positions of the pulsars. 

Throughout the paper the scale radius of the King model, $r_0$, will be considered equal to the core radius, $r_c$, defined as the radius at which the projected luminosity density falls to half its central value. The line-of-sight positions $l$ will be measured from the centre of the cluster increasing away from the observer.

\subsection{Positions}

The column density profile of the pulsars in a globular cluster following a King profile can be well approximated within a few core radii with the formula from \cite{Lugger1995}:

\begin{equation} \label{dens_perp}
n(x_{\perp})= n_0 \left(1 + x_{\perp}^2 \right)^{\alpha/2},
\end{equation}
where $n_0$ is the central density, and $x_{\perp}$ is the distance from the centre in the plane of the sky in units of core radii, defined as $x_{\perp}=R_{\perp}/r_c$. The power law index  $\alpha$ is linked to mass segregation and it is related to the mass of the pulsars by the relation $\alpha= 1- 3q$, where $q$ is the ratio between the mass of the pulsar and the dominant mass class of the cluster ($q= \rm{M_{p}/M_{*}}$). In the case of pulsars having the same mass as the dominant mass class we recover $\alpha=-2$, which is the value for the single-mass analytical King model \citep{King1962}. 

The three-dimensional number density has been calculated by \cite{Grindlay2002} and is:

\begin{equation} \label{dens_3d}
n(x) \propto \left( 1 + x^2 \right)^{(\alpha-1)/2},
\end{equation}
where $x=r/r_c$ and $r$ is the three-dimensional position of the pulsar.

Also in this case, if $\alpha=-2$, we recover the spatial density profile of the single mass King model.

\subsection{Velocity distribution}\label{section_velocity_distr}

The average square velocity of stars in a globular cluster can be obtained from the King distribution function, which is defined as follows:
\begin{equation}
f_K(\mathcal{E})= \begin{cases} \rho_1 (2\pi \sigma_{\rm vel}^2)^{-3/2} \,({\rm e}^{\mathcal{E}/\sigma_{\rm vel}^2}-1)  & \mathcal{E}>0 \\ 0 & \mathcal{E}\leq 0 \end{cases},
\end{equation}
where $\mathcal{E}$ is the relative energy defined as $\mathcal{E} = \Psi -\frac{1}{2}v^2$, and $\Psi$ is the gravitational potential energy, $\rho_1$ is a reference density, $\sigma_{\rm vel}$ is the central one-dimensional velocity dispersion and $v$ is the three-dimensional velocity of the star.

The average square velocity can be recovered by the integral:

\begin{equation}
\langle v^2\rangle = \frac{\int_{0}^{\infty} v^4 f_K(\mathcal{E})dv}{\int_{0}^{\infty} v^2 f_K(\mathcal{E})dv}.
\end{equation}

The limitation of $\mathcal{E}>0$ can be implemented by limiting the integrals from 0 to $\sqrt{2\Psi}$. In this way we find the solution:

\begin{equation}
\langle v^2\rangle = \frac{3\sigma_{\rm vel}^2 e^{\Psi/\sigma_{\rm vel}^2} \rm{erf}\left(\sqrt{\frac{\Psi}{\sigma_{\rm vel}^2}}\right) -\frac{6}{\sqrt{\pi}} \sqrt{\Psi} \sigma_{\rm vel} -\frac{4}{\sqrt{\pi} \sigma_{\rm vel}} \Psi^{3/2}- \frac{8}{5\sqrt{\pi}} \frac{\Psi^{5/2}}{\sigma_{\rm vel}^3}  }{ e^{\Psi/\sigma_{\rm vel}^2} \rm{erf}\left(\sqrt{\frac{\Psi}{\sigma_{\rm vel}^2}}\right) - \sqrt{\frac{4\Psi}{\pi \sigma_{\rm vel}^2}} \left(1 +\frac{2\Psi}{3\sigma_{\rm vel}^2}\right) },
\end{equation} 
where erf is the Gauss error function.

This expression can be approximated within 20 core radii and with a maximum error of 2 per cent by the formula:

\begin{equation} \label{average_sqrvel}
\langle v^2\rangle = \sqrt{3} \sigma_{\rm vel} \left[ 1+ \left(\frac{x}{6}\right)^2\right]^{-0.2}
\end{equation}

This velocity distribution is valid for the dominant mass class of the cluster. The mass of the dominant class is close to the main sequence turn-off that is $\sim 0.8 \, \rm{M_{\odot}}$ while pulsars are typically more massive weighing around $1.4 \, \rm{M_{\odot}}$. Since globular clusters evolve towards energy equipartition, we should expect the pulsars to have lower velocities. \cite{Bianchini2016} estimated that the true equipartition is reached only for stars whose mass M is above a certain equipartition mass. The velocity dispersion for each mass is:
\begin{equation}
\sigma_{\rm vel}(M)= 
\begin{cases} \sigma_{\rm vel} \exp\left(-\frac{1}{2} \rm{\frac{M}{M_{eq}}}\right) & \text{if } \rm{M\leq M_{eq}} \\ \sigma_{\rm vel, eq}\left( \rm{\frac{M}{M_{eq}}}\right)^{-1/2}  & \text{if } \rm{M> M_{eq}}\end{cases}
\end{equation}
where $\sigma_{\rm vel, eq} = \sigma_{\rm vel}\exp(-\frac{1}{2})$ is the velocity dispersion at the equipartition mass.

For 47 Tuc, \cite{Baldwin2016} measured the equipartition mass to be 1.6 M$_{\odot}$. So for the pulsars, assuming a mass of $\sim 1.4$ M$_{\odot}$, we obtain that the central velocity dispersion is $\sigma_{\rm{vel, pulsar}} \sim 0.65 \sigma_{\rm vel}$. This is the value that must be used in equation (\ref{average_sqrvel}).

\subsection{Acceleration}

The acceleration acting on a pulsar inside a globular cluster is due both to the gravitational potential as modelled by the King profile and to the perturbations caused by the nearby stars. \cite{Prager2017} showed in Terzan 5 that the acceleration from the nearest neighbours is negligible if compared to the mean field acceleration. The same is considered to be valid also for 47 Tuc. 
The acceleration for the King profile was derived explicitly by \cite{Freire2005} and, within a few core radii, takes the value:

\begin{equation}
a_r(x)= - 4\pi G \rho_c \theta_c x^{-2} d\left[ \rm{arcsinh}(x) - \frac{x}{\sqrt{1+x^2}} \right],
\end{equation}
where $\rho_c$ is the central density, $\theta_c$ is the angular core radius, $d$ is the distance to the cluster.
Projecting this acceleration along the line of sight we obtain:

\begin{equation} \label{acc_lpos}
a_r(l,x)= - 4\pi G \rho_c x^{-3} l \left[ \rm{arcsinh} (x) - \frac{x}{\sqrt{1+x^2}} \right],
\end{equation}
where $l$ is the line-of-sight component of the position of the pulsar relative to the centre of the cluster, in core radii.

For a given position in the plane of the sky, $x_{\perp}$, the acceleration has a maximum value determined numerically for each line of sight, at the centre ($x_{\perp} = 0$) this is given by \citep{Freire2017}:
\begin{equation}
a_{\rm l, max}(x_{\perp}) =1.5689\frac{\sigma_{\mu,0}^2 d}{\theta_c}.
\end{equation}

The proper motion central velocity dispersion, $\sigma_{\mu,0}$, is defined as in equation \ref{vel_disp_eq} and is related to the one-dimensional velocity dispersion defined in Section \ref{section_velocity_distr}, $\sigma_{\rm vel}$, by the equation: $\sigma_{\rm vel}=\sigma_{\mu,0}\, d$.

\begin{figure}
\begin{center}
\includegraphics[width=\columnwidth]{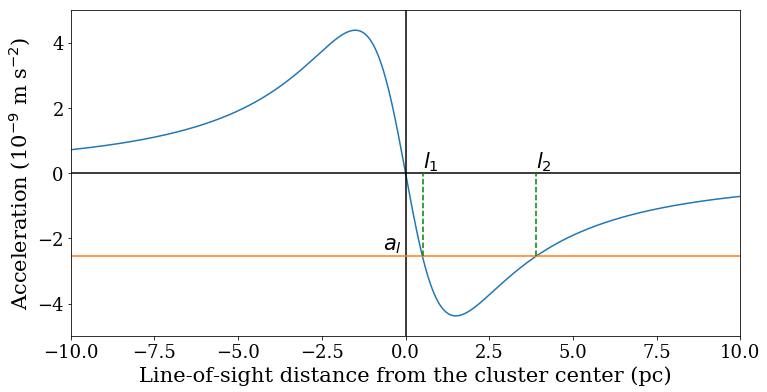}
\caption{ Plot of the acceleration along the line of sight as a function of the line-of-sight distance from the cluster centre. This plot has been derived for pulsar 47 Tuc C. $a_l$ is the measured acceleration for the pulsar in consideration. To generate this acceleration the pulsar could be located either at $l_1$ or in $l_2$.
}\label{degeneracy}
\end{center}
\end{figure}

The shape of the acceleration along the line of sight is shown in Figure \ref{degeneracy}. For a given acceleration $a_l$ there are two possible line-of-sight positions that are compatible. Therefore using only the measurement of the acceleration it is not possible to determine unequivocally the position of the pulsar.

\subsection{Measuring accelerations for binary pulsars}\label{binary_pulsars}

The value of the acceleration along the line of sight at the pulsar position can be recovered by looking at the derivative of the rotational period. The measured period derivative can be expressed as:

\begin{equation} \label{pdotp_meas}
\left( \frac{\dot P}{P}\right)_{\rm meas} = \left( \frac{\dot P}{P}\right)_{\rm int} + \frac{a_c}{c} +\frac{a_g}{c} +\frac{\mu^2 d}{c},
\end{equation}
where $\left( \frac{\dot P}{P}\right)_{\rm int}$ is the intrinsic spin-down of the pulsar, $ a_c/c$ is the acceleration due to the gravitational potential of the cluster, $a_g/c$ is the acceleration due to the Galactic potential and $\mu^2 d/c$ is the Shklovskii effect \citep{Shklovskii1970} and $\mu$ is the proper motion of the pulsar. To obtain the acceleration due to cluster gravitational potential we first need to estimate the other contributions. 

The hardest contribution to determine is the intrinsic spin-down of the pulsar as this quantity cannot be measured directly. In the case of binary pulsars, however, we can also measure the orbital period and, in some cases, the orbital period derivative. Since the orbital period is subjected to the same effects as the rotational period, eq. (\ref{pdotp_meas}) applies also for this quantity. 
Gravitational wave emission is the dominant source of the intrinsic orbital period derivative in most of the binary systems present in the cluster.
As shown in \cite{Freire2017} the effects of the gravitational wave emission are much weaker than that of the cluster potential. Therefore:

\begin{equation}
\left( \frac{\dot P_b}{P_b}\right)_{\rm meas} = \frac{a_c}{c} +\frac{a_g}{c} +\frac{\mu^2 d}{c}.
\end{equation}

Since we can measure the Galactic potential acceleration and the Shklovskii effect, a measurement of the orbital period derivative is a direct measurement of the acceleration due to the cluster potential.

The acceleration from the Galactic potential can be measured starting from the distance to the Galactic Centre, taken to be $R_0=8.34 \pm 0.16$ kpc \citep{Reid2014} and from the rotational speed of the Galaxy at the Sun position, $\Theta_0=240 \pm 8$ km s$^{-1}$ \citep{Sharma2014}. The acceleration due to the differential Galactic rotation at the distance of 47 Tuc, $d=4.53\pm 0.08$ kpc \citep{Bogdanov2016} and its location given in Galactic coordinates ($l=305.8953$, $b=-44.8891$) is \citep{Nice1995}:

\begin{equation}
a_g= -\cos(b_{\rm gal}) \left( \frac{\Theta_0^2}{R_0}\right) \left[ \cos(l_{\rm gal}) + \frac{\beta}{\sin^2(l_{\rm gal})+\beta^2}\right]\;\;\rm{m\, s}^{-2},
\end{equation}
where $l_{\rm gal}$ and $b_{\rm gal}$ are the Galactic coordinates and $\beta= (d/R_0)\cos(b_{\rm gal})-\cos(l_{\rm gal})$. We obtain $a_g \simeq -5\times 10^{-11}$ m s$^{-2}$ which is two orders of magnitude smaller then the measured accelerations.

The acceleration due to the Shklovskii effect can be measured directly from the proper motion of the pulsars.
For the large majority of pulsars in 47 Tuc, the proper motion has been measured and the Shklovskii effect can be calculated. The average acceleration due to the Shklovskii effect can be estimated by using the proper motion of the cluster measured with data from Gaia Data Release 2. The measured proper motion is $\mu_{\rm RA}=5.2477 \pm 0.0016$ mas yr$^{-1}$ and $\mu_{\rm DEC}=-2.5189 \pm 0.0015$ mas yr$^{-1}$ \citep{Gaia2018}.
This average acceleration will therefore be $\simeq 1\times10^{-10}$ m s$^{-2}$. 

Among the binary pulsars known in 47 Tuc there are also four black widow systems. The long-term timing of some black widows \citep{Shaifullah2016} shows very large and unpredictable variability of the orbital period. In this case it is not possible to estimate the orbital period derivative. \cite{Freire2017} showed that two black widow systems, 47 Tuc J and O, show such large orbital variability. However, the other two, 47 Tuc I and R, show small orbital variability and their period derivative can be described as due to the cluster acceleration.

\subsection{Measuring accelerations for isolated pulsars} \label{acceleration_isolated}

For isolated pulsars and for the binaries for which we cannot measure the orbital period derivative, we need a way to estimate the intrinsic spin-down. The intrinsic spin-down can be linked to the surface magnetic field of the pulsar so if we assume that these MSPs have similar properties to the ones found in the Galactic field, then we can estimate the surface magnetic field. A list of all the pulsars of interest can be found in the Australian Telescope National Facility (ATNF) Pulsar Catalogue\footnote{The full list is available at \href{http://www.atnf.csiro.au/research/pulsar/psrcat/}{http://www.atnf.csiro.au/research/pulsar/psrcat/}}  \citep{Manchester2005}. \cite{Prager2017} found that the surface magnetic field values of field MSPs can be fitted with a log-normal probability distribution function with $\mu_{\log_{10}(B)}= 8.47$ and $\sigma_{\log_{10}(B)}= 0.33$.

The average apparent acceleration caused by  the intrinsic spin-down is $\sim 1 \times 10^{-9}$ m s$^{-2}$. This is a significant contribution when compared to the range of accelerations due to the globular cluster as can be seen in Figure \ref{degeneracy}.

\subsection{Jerks}\label{section_jerks}

The effects of the jerks can be measured from the second-order derivative of the observed period of the pulsar, as described by \cite{Phinney1993}:

\begin{equation}
\frac{\ddot P}{P} = \frac{1}{c}  \dot{\bf{a}} \cdot \bf{n},
\end{equation}
where $\bf{n}$ is the direction of the line of sight.

Also, we need to distinguish between the jerk arising from the mean gravitational potential and the jerk caused by the nearest neighbours. However, in this case, the two contributions are of similar magnitude and must be considered together \citep{Prager2017}. The jerk due to the mean field was calculated by \cite{Phinney1993} to be:

\begin{equation}\label{jerk_mean_field}
 \dot{\bf{a}}_{\rm{mf}} \cdot {\bf{n}} = \frac{4}{3} \pi G \rho(r) v_l(r),
\end{equation}
where $v_l(r)$ is the line-of-sight component of the velocity of the pulsar. A comparison between the maximum jerk caused by the mean field and the measured jerk for the pulsars in 47 Tuc was done by \cite{Freire2017}.

The jerk caused by the nearest neighbours follows a Lorentzian distribution with scale parameter given by \cite{Prager2017}:

\begin{equation}\label{jerk_nearest_neighbour}
\dot a_{nn} = \frac{2\pi \xi}{3} G \langle m \rangle \sigma_{\rm vel} n,
\end{equation}
where $\xi \simeq 3.04$ is a numerical factor, $\langle m \rangle$ is the average mass of the neighbouring stars, $\sigma_{\rm vel}$ is the velocity dispersion and $n$ is the number density of the stars near the pulsar.

\subsection{Effects of an IMBH}

The presence of an IMBH in the centre of the cluster will have an influence on the density distribution of the stars inside a certain radius, called the influence radius and defined as \citep{Baumgardt2004a}:

\begin{equation}
r_i =\frac{3 \rm{M_{BH}}}{8\pi \rho_c r_c^2},
\end{equation} 
where M$_{\rm BH}$ is the mass of the central black hole.

Outside this radius the distribution is the same as the standard King model described above. Inside, the dynamics is dominated by the black hole and the stars follow a density cusp described by the power law distribution:

\begin{equation}
\rho_{\rm BH} \propto r^{-1.55},
\end{equation}
where the index of the power law is taken from \cite{Baumgardt2004b}.

This difference in density will cause a difference in the acceleration felt by the pulsars inside or close to the influence radius of a given black hole. 

The equations described above are valid only if the black hole is taken as fixed at the centre of the cluster. While this is a good assumption if the IMBH is massive, a black hole of smaller mass may wander in the central region disrupting the central density cusp. Therefore this test is valid only to determine an upper mass limit on a putative IMBH.

\section{MCMC analysis}\label{section_mcmc}

For the analysis of the gas distribution inside a globular cluster we need to accurately know the three-dimensional position of each pulsar within the cluster itself and compare the predicted DM (in turn dependent on the adopted distribution model for the gas) with the measured DM. The position along the line of sight can be estimated using the formulas shown above. 
Other parameters, like the core radius, the central density, the density distribution power law index and the line-of-sight velocities are needed to obtain these results. 
All of these parameters are left free. 
Since all of these parameters, except for the line-of-sight velocities, are well constrained for this cluster, the agreement of the estimated values with those presented in the literature is an indicator of the quality of the fit. 
The distance of the cluster is not taken as a parameter of the fit but is taken as fixed. 
Trying to consider the distance as a free parameter only resulted in inconclusive fits.

The data we use to perform the fit are: the rotational period, the rotational period derivative, the second-order rotational period derivative, the orbital period, the orbital period derivative, the positions of the pulsars projected along the plane of the sky and their proper motions. Since the formulas are valid only near the cluster centre we cannot use the information about pulsar 47 Tuc X, which is about 5 pc away. The total number of pulsars we can use for the fit is thus 22 (pulsars P and V do not have a phase-connected solution yet). Pulsar H exhibits a very large jerk, which has been suggested to be caused by nearby stars, so we cannot use this measurement for our fit. This brings the total number of parameters of the fit to 46. If we also search for the presence of an IMBH we must also fit for the mass of the black hole, which brings the total number of parameters to fit to 47.

The analysis was performed using the {\sc emcee} python package \citep{Foreman-Mackey2013} which implements a Markov Chain Monte Carlo (MCMC) algorithm and returns the best fit parameters for the desired model. 

\subsection{Likelihoods}

The MCMC algorithm works by looking for the set of parameters that maximises the likelihood. The likelihood passed to the algorithm is expressed in logarithms and can be seen as the sum of different log-likelihoods:

\begin{equation}
\mathcal{L} = \mathcal{L}_{\rm x_{\perp}} +\mathcal{L}_{\rm l}+ \mathcal{L}_{\rm accel} + \mathcal{L}_{\rm jerks} +\mathcal{L}_{\rm vel},
\end{equation}
where $\mathcal{L}_{\rm x_{\perp}}$ is the log-likelihood associated with the pulsar position in the plane of the sky, $\mathcal{L}_{\rm l}$ is the log-likelihood associated to their three-dimensional position, $\mathcal{L}_{\rm accel}$ is the log-likelihood due to the experienced acceleration,     $\mathcal{L}_{\rm jerks}$ is the log-likelihood due to the jerk measurements and $\mathcal{L}_{\rm vel}$ is the log-likelihood associated with the velocity measurements.

The log-likelihood associated with the position of the pulsars on the plane of the sky can be found starting from the number density distribution of this stellar component on the plane of the sky (eq. \ref{dens_perp}):

\begin{equation}
\mathcal{L}_{\rm x_{\perp}} \propto \sum_i \log_{10} \left[ \left( 1 +x_{\perp,i}^2 \right)^{\alpha/2} \right],
\end{equation}
where $i$ is the index of the summation over all pulsars.

The log-likelihood associated with the three-dimensional position of the pulsars in the cluster is (eq. \ref{dens_3d}): 

\begin{equation}
\mathcal{L}_{\rm l} \propto \sum_i \log_{10} \left[ \left( 1 +x_{\perp,i}^2 +\frac{l_i^2}{r_{\rm c}^2}\right)^{(\alpha-1)/2} \right].
\end{equation}

The acceleration log-likelihood is measured in two different ways depending on whether the pulsar is in a binary system with a measured orbital period derivative. If we know the latter, we can directly probe the acceleration and compare it against the one predicted by the model (eq. \ref{acc_lpos}). The log-likelihood then becomes:

\begin{equation}
\mathcal{L}_{\rm acc, binary} \propto \sum_i \frac{1}{2\epsilon_i} (a_{l,i}- a(l | x_{\perp}, \theta))^2,
\end{equation}
where $\epsilon_i$ is the uncertainty on the measured acceleration, $a_{l,i}$ is the measured acceleration and $a(l | R_{\perp}, \theta)$ is the predicted acceleration for the set of parameters $\theta$.

If the pulsar is isolated or we have no measurement of the orbital period derivative, we have to estimate the intrinsic spin-down due to magnetic braking. We first subtract the model acceleration from the measured $(\dot P/P)$ and then check if the residual acceleration could be due to the intrinsic spin-down. As described in Section \ref{acceleration_isolated}, this quantity can be linked to the surface magnetic field of the pulsar. The magnetic fields of Galactic MSPs follow a log-normal distribution and the log-likelihood becomes:

\begin{equation}
\mathcal{L}_{\rm acc, isolated} \propto \sum_i \left[ \frac{1}{2\sigma_{\log_{10}(B)}^2}(\log_{10} B_8-\mu_{\log_{10}(B)})^2 + \log_{10} B_8 \right],
\end{equation}
where $B_8$ is the magnetic field in units of $10^8$ G, whereas $\sigma_{\log_{10}(B)}$ and $\mu_{\log_{10}(B)}$ are the parameters of the lognormal fit performed by \cite{Prager2017} on the Galactic MSPs as described in Section \ref{acceleration_isolated}.

As shown in Section \ref{section_jerks}, the jerk to which a pulsar is subject is due to both the mean field potential and to nearby stars. The jerk due to the mean field can be estimated directly from the formulas while for the stellar contribution only a statistical description is possible. To estimate the likelihood of measuring a certain value for the jerk we subtract the mean field component and compute the logarithm of the probability that the residual is caused by nearby stars. The log-likelihood becomes:

\begin{equation}
\mathcal{L}_{\rm jerks} = \sum_i \log_{10} \left( \frac{\dot{a}_{nn,i}}{\pi} \frac{1}{[(\dot{a}_{l,i} - \dot{a}_{\rm mf,i})^2 + \dot{a}_{nn,i}^2]}\right), 
\end{equation}
where $\dot{a}_{l,i}$ is the measured jerk, $\dot{a}_{\rm mf,i}$ is the predicted mean field jerk and $\dot{a}_{nn,i}$ is the scale parameter of the Lorentzian distribution (eq. \ref{jerk_nearest_neighbour}).

The velocities of the pulsars are distributed according to a Maxwellian distribution, with a velocity dispersion which can be estimated for each pulsar from equation (\ref{average_sqrvel}). Hence the log-likelihood for the velocity is:

\begin{equation}
\mathcal{L}_{\rm vel} \propto \sum_i \left( -3 \log_{10}(\langle v^2\rangle_i) + 2\log_{10}(v_{\rm meas, i}) - \frac{3}{2}\frac{v_{\rm meas, i}^2}{\langle v^2\rangle_i^2}\right).
\end{equation}

\subsection{Priors}

Priors were initially chosen to be flat for all parameters except for the black hole mass, which can range by orders of magnitude; therefore a logarithmic prior is more reasonable. However, since the fit did not converge we decided to put a gaussian prior on the core radius centred in 0.58 pc and with a standard deviation of 0.03 pc, as derived from the recent optical study of \cite{Bellini2017}.

\subsection{Parallel tempering}

As shown in Fig. \ref{degeneracy} it is possible for two different positions along a given line of sight to produce the same line-of-sight acceleration. This generates a bimodal distribution of the line-of-sight position of a pulsar for every measured acceleration. Since the MCMC could get stuck on one of the two solutions and not explore the parameter space properly, we need to address this problem. We opted for a parallel tempering solution \citep{Marinari1992} which makes use of chains of different `temperatures' to cover the entire parameter space. The `higher temperature chains are allowed to move freely while the `colder' chains remain close to the previous values. Combining chains of different temperatures allows us to properly explore the parameter space in order to find the global maximum of the likelihood.

\subsection{Fit results}

\begin{figure}
\begin{center}
\includegraphics[width=\columnwidth]{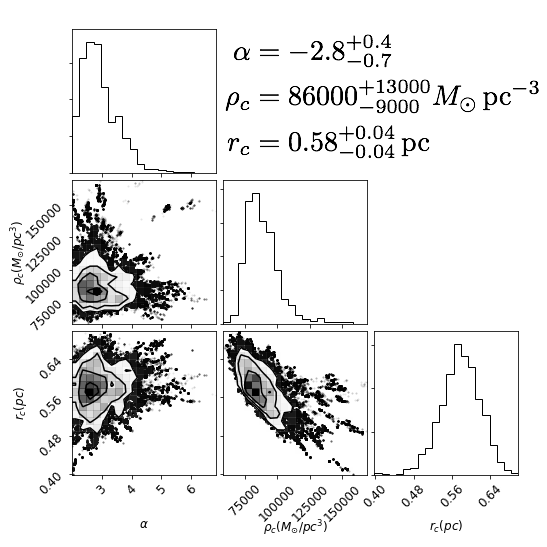}
\caption{Triangle plot showing the marginalized probabilities for the  density distribution power law index $\alpha$, the central density, $\rho_c$, and the core radius, $r_c$, for the globular cluster 47 Tuc. The errors indicate the 68\% credible intervals.
}\label{triangle_jerk}
\end{center}
\end{figure}

The best fit for parameters of 47 Tuc are shown in Fig. \ref{triangle_jerk}. The results of the fit for the position along the line of sight are reported in Table \ref{MCMC_results_table}.

The posterior distribution of the core radius does not show asymmetries or deviations from the assumed Gaussian prior. This means that the fit is not strongly influenced by this parameter. Instead, the best fit result is the value we assumed for the prior. 

\begin{figure}
\begin{center}
\includegraphics[width=\columnwidth]{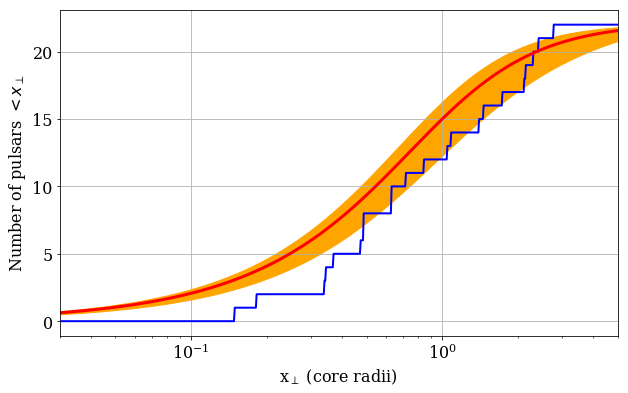}
\caption{Cumulative distribution of the projected offset from the cluster centre. The red line has been obtained integrating eq. (\ref{dens_perp}) using the parameter $\alpha$ found in the MCMC fit. The orange area is the 68\% credible interval.
}\label{Cumulative}
\end{center}
\end{figure}

The power law index of the density distribution, $\alpha$, is found to be $-2.8_{-0.7}^{+0.4}$. The errors indicate the 68\% credible interval of the posterior distribution. This value is consistent with the value of $-3.26\pm 0.48$ measured in X-rays for the MSPs  of 47 Tuc\citep{Heinke2005}. To check the consistency we also compare, in Fig. \ref{Cumulative}, the cumulative distribution of the pulsars in the plane of the sky with the result obtained with eq. (\ref{dens_perp})  using the derived power law index.

The best-fit value for the central density is $\rho_c =  8.6 ^{+ 1.3}_{- 0.9}\times 10^4 \, \rm{M}_{\odot}\, \rm{pc}^{-3}$. This value is compatible with the previously estimated central density of $(7.5 \pm 0.2) \times 10^4 \, \rm{M}_{\odot}\, \rm{pc}^{-3}$ (Table~\ref{tab47tuc}).

\begin{figure}
\begin{center}
\includegraphics[width=\columnwidth]{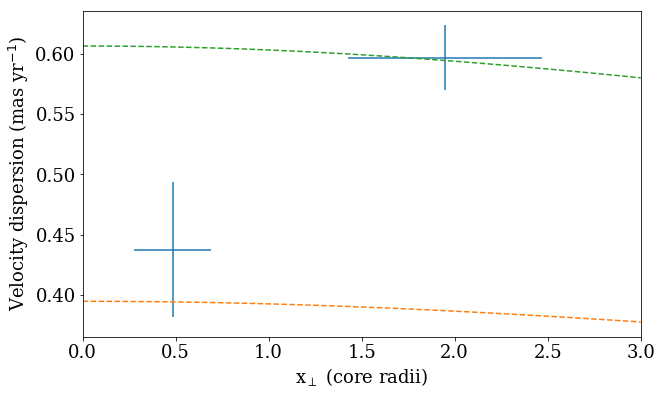}
\caption{One-dimensional velocity dispersion measured from the proper motion motion of the pulsars as a function of radius. The pulsars have been divided in two sets, one inside the core radius and one outside. The error bars show the 1-$\sigma$ interval. The green curve shows the predicted trend for stars of the same mass as the dominant mass class while the orange line shows the predicted trend for stars of $\sim 1.4$ M$_{\odot}$. The two curves are described in Section \ref{section_velocity_distr}.
}\label{velocity_distr}
\end{center}
\end{figure}

From the estimates of the core radius and central density we can calculate the velocity dispersion through equation \ref{vel_disp_eq}. We obtain $\sigma_{\mu,0}= 0.60 \pm 0.04$ mas yr$^{-1}$ which is compatible with velocity dispersion measured from the proper motion of the stars \citep{Watkins2015}.
The calculated velocity dispersion can also be compared with the pulsar proper motions thanks to the equations described in Section \ref{section_velocity_distr}. The result is shown in Figure \ref{velocity_distr} where the velocity dispersion is measured separately for the pulsars inside the core radius and those outside. The pulsars closer to the centre have a velocity dispersion close to what is expected for their mass. The pulsars outside the core radius have higher velocity dispersions than expected probably because they have not reached energy equipartition.

\begin{figure}
\begin{center}
\includegraphics[width=\columnwidth]{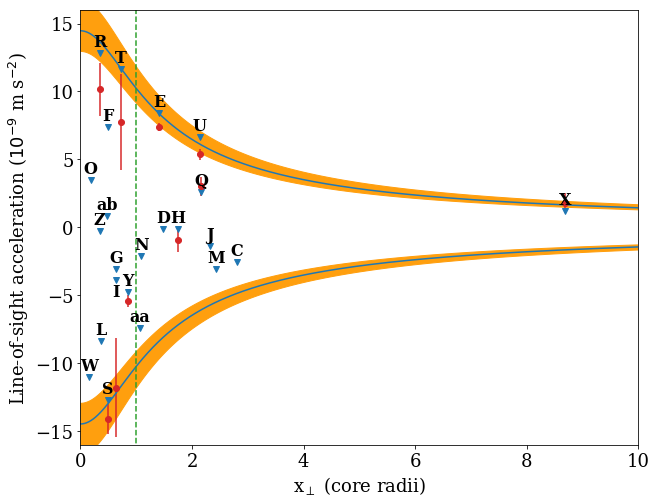}
\includegraphics[width=\columnwidth]{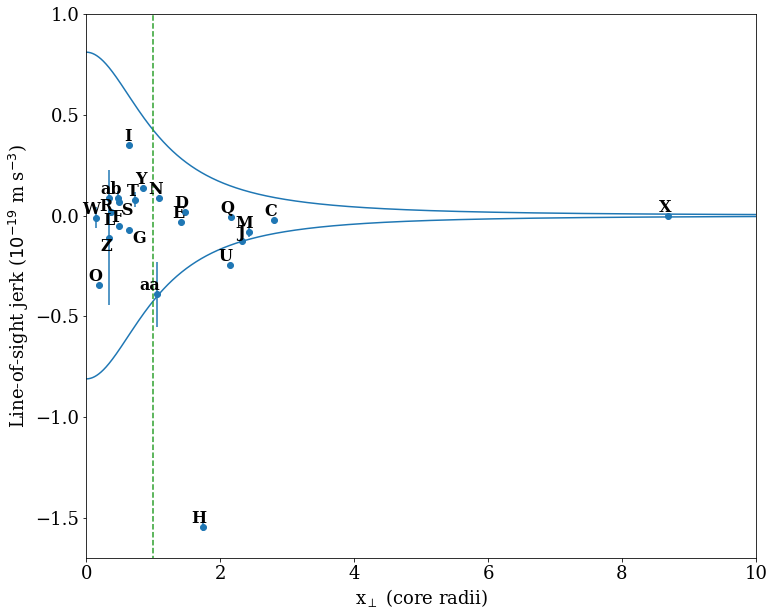}
\caption{The top panel shows the line-of-sight acceleration of the pulsars as a function of the offset from the centre in units of core radii. The blue triangle shows the measure of $(\dot P/P)_{\rm meas}$ which is an upper limit on the true value of the acceleration. For pulsars in binaries for which a measure of the orbital period derivative is available we show the measured acceleration (for a discussion on this see Sections \ref{binary_pulsars} and \ref{acceleration_isolated}). The blue curve shows the maximum and minimum acceleration in absence of a black hole and the orange area is the 68\% credible interval. 
The bottom panel shows the line-of-sight jerks as a function of the offset from the centre in units of core radii. The blue curves show the maximum and minimum jerks due to the mean field in absence of a black hole. In both panels the vertical dotted green line is the core radius.
}\label{acc_jerk_plots}
\end{center}
\end{figure}

Since we measured new values for the structural parameters of 47 Tuc, we can also check the plots of the line-of-sight accelerations and jerks. These plots are shown in Fig. \ref{acc_jerk_plots}. When compared with the similar plots presented in \cite{Freire2017}, the main difference is the acceleration of the pulsar S which now is below the minimum possible acceleration in absence of the black hole. It is however still compatible with the limit considering the large errors on the structural parameters of the cluster.
Therefore, all the main structural parameters estimated for the cluster are compatible with those previously measured. This gives confidence in the reliability of the algorithm used and in the line-of-sight positions presented in Table \ref{MCMC_results_table}.

In all models we assumed that the cluster was spherically symmetric. To verify whether this assumption is consistent with the results, we perform a Kolmogorov-Smirnoff test to check if the measured positions along the line of sight are extracted from the same distribution as the positions along two directions on the plane of the sky. Both tests with right ascension and declination return p-values of $\sim 0.6$ so the results are consistent with a spherically symmetric cluster.

\begin{table}
\caption{Best-fit parameters for the line-of-sight position of the pulsars. The distance from the centre of the cluster along the plane of the sky, $R_{\perp}$, is also shown together with the DM values. The errors indicate the 68\% credible interval on the posterior distribution.}

\def\mystrut{\rule{0pt}{1\normalbaselineskip}}
\begin{center}
\begin{tabular}[t]{c c c c }
\hline
{Pulsar} & {$R_{\perp}$ (pc)} & {DM (pc cm$^{-3})$} & {l (pc)}  \\
\hline
\hline
W & $ 0.087 $ & $ 24.4 \pm 0.5 $      & $0.40 _{-0.06}^{+0.29}$ \\ \mystrut
O & $ 0.106 $ & $ 24.356 \pm 0.002 $  & $-0.03 _{-0.03}^{+0.16}$ \\\mystrut 
Z & $ 0.198 $ & $ 24.4 \pm 0.5 $      & $0.01 _{-0.00}^{+0.08}$ \\ \mystrut
R & $ 0.200 $ & $ 24.361 \pm 0.007 $  & $-0.22 _{-0.15}^{+0.08}$ \\\mystrut
L & $ 0.214 $ & $ 24.400 \pm 0.012 $  & $0.24 _{-0.07}^{+0.26}$ \\ \mystrut
ab & $ 0.276 $ & $ 24.373 \pm 0.020 $ &  $0.02 _{-0.03}^{+0.12}$ \\\mystrut
F & $ 0.283 $ & $ 24.382 \pm 0.005 $  & $-0.11 _{-0.07}^{+0.22}$ \\\mystrut
S & $ 0.283 $ & $ 24.376 \pm 0.004 $  & $0.58 _{-0.19}^{+0.17}$ \\ \mystrut
I & $ 0.365 $ & $ 24.429 \pm 0.010 $  & $0.33 _{-0.22}^{+0.15}$ \\ \mystrut
G & $ 0.367 $ & $ 24.436 \pm 0.004 $  & $0.11 _{-0.03}^{+0.16}$ \\ \mystrut
T & $ 0.419 $ & $ 24.411 \pm 0.021 $  & $-0.28 _{-0.22}^{+0.17}$ \\\mystrut
Y & $ 0.493 $ & $ 24.468 \pm 0.004 $  & $0.19 _{-0.04}^{+0.03}$ \\ \mystrut
aa & $ 0.613 $ & $ 24.971 \pm 0.007 $ &  $0.62 _{-0.18}^{+0.23}$ \\\mystrut
N & $ 0.631 $ & $ 24.574 \pm 0.009 $  & $0.17 _{-0.07}^{+0.28}$ \\ \mystrut
E & $ 0.818 $ & $ 24.236 \pm 0.002 $  & $-0.54 _{-0.12}^{+0.14}$ \\\mystrut
D & $ 0.854 $ & $ 24.732 \pm 0.003 $  & $0.04 _{-0.02}^{+0.11}$ \\ \mystrut
H & $ 1.012 $ & $ 24.369 \pm 0.008 $  & $0.10 _{-0.07}^{+0.05}$ \\ \mystrut
U & $ 1.237 $ & $ 24.337 \pm 0.004 $  & $-0.80 _{-0.19}^{+0.26}$ \\\mystrut
Q & $ 1.252 $ & $ 24.265 \pm 0.004 $  & $-0.33 _{-0.12}^{+0.13}$ \\\mystrut
J & $ 1.342 $ & $ 24.588 \pm 0.003 $  & $0.54 _{-0.25}^{+0.40}$ \\ \mystrut
M & $ 1.409 $ & $ 24.432 \pm 0.016 $  & $0.66 _{-0.19}^{+0.37}$ \\ \mystrut
C & $ 1.620 $ & $ 24.600 \pm 0.004 $  & $0.75 _{-0.28}^{+0.78}$ \\
\hline

\end{tabular}\\
\label{MCMC_results_table}

\end{center}

\end{table}

The line of sight velocities are not tightly constrained by our code. They are  considered nuisance parameters over which we marginalize. As a result we obtain values of line of sight velocities with very large uncertainties. 

\begin{figure}
\begin{center}
\includegraphics[width=\columnwidth]{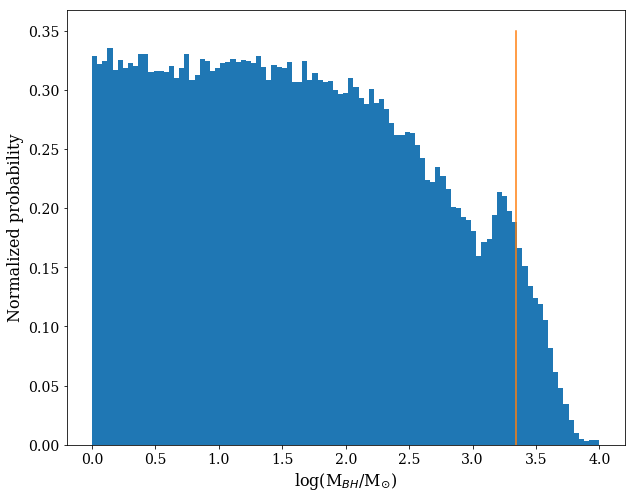}
\caption{Plot of the posterior probability on the mass of the central black hole. The maximum of the distribution is at 0 M$_{\odot}$ suggesting that a black hole is not necessary to explain the data. The vertical line is at the mass of the black hole of $2200 \rm{M}_{\odot}$ proposed by \protect\cite{Kiziltan2017a}. We see an increase in probability close to this value. The upper mass limit above which the black hole is not compatible with our data is $\sim 4000 \rm{M}_{\odot}$.
}\label{IMBH_posterior}
\end{center}
\end{figure}

Furthermore the posterior distribution of the IMBH mass is shown in Fig. \ref{IMBH_posterior}. The peak of the distribution is at zero mass, suggesting that there is no IMBH at the cluster centre. We can derive an upper limit on the mass by measuring the value that contain 99\% of the chains. This limit is at $\sim 4000 \,\rm{M}_{\odot}$.

\section{Gas model results}\label{section_gas} 

By using the information (derived in the previous sections) about the positions along the line of sight of the pulsars it is possible to investigate which model for the distribution of the internal gas better matches the observed DMs of the 47 Tuc pulsars. We performed a fit to the data presented in Table \ref{MCMC_results_table} with a Bayesian algorithm without considering pulsars W and Z, because of their imprecise DMs. That allows us to compare models with different parameters through the Bayes factor. 
This consists of measuring the evidences, which in Bayesian statistics is the integral along the entire parameter space of the likelihood. The evidences can then be compared by calculating the ratio. If the logarithm in base 10 of this ratio is larger than 2, the model with the highest evidence is strongly favoured.

The line-of-sight positions of the pulsars are not normally distributed and  therefore standard fitting procedures would not work. To correctly treat their distribution we extracted the values of the line-of-sight positions at each cycle of the algorithm from the posterior distributions, and built the uncertainty range including the 68\% of the posterior density function.

From a given gas density distribution, the contribution $\rm{DM_{GC}}$ to the total observed ${\rm DM}$  due to the globular cluster can be measured with the following integral for each pulsar:

\begin{equation}
{\rm DM_{GC}}= \int_{-l_T}^{l} n_g(R_{\perp}, l') dl' ,
\end{equation}
where $l_T$ is the tidal radius of the cluster, which is assumed as the maximum radius up to which the gas is present.

\subsection{Constant density model}

We first tested the hypothesis of a constant gas density in the region of interest. This is the model that was used by \cite{Freire2001b} to give the first evidence of ionized gas inside the cluster. Assuming that the region of interest is uniformly permeated by a gas, we do not need to consider the position in the plane of the sky. The total DM for each pulsar is described by the formula:

\begin{equation}
{\rm DM}= n_g  l + \rm{DM_c},
\end{equation}
where $n_g$ is the value of the gas density and $\rm{DM_c}$ is the value of the total $\rm{DM}$ at  a plane that passes through the centre of the cluster and is perpendicular to the line of sight (assuming no variation of $\rm{DM}$ due to the interstellar medium along the various lines of sight to the globular cluster). 

\begin{figure}
\begin{center}
\includegraphics[width=\columnwidth]{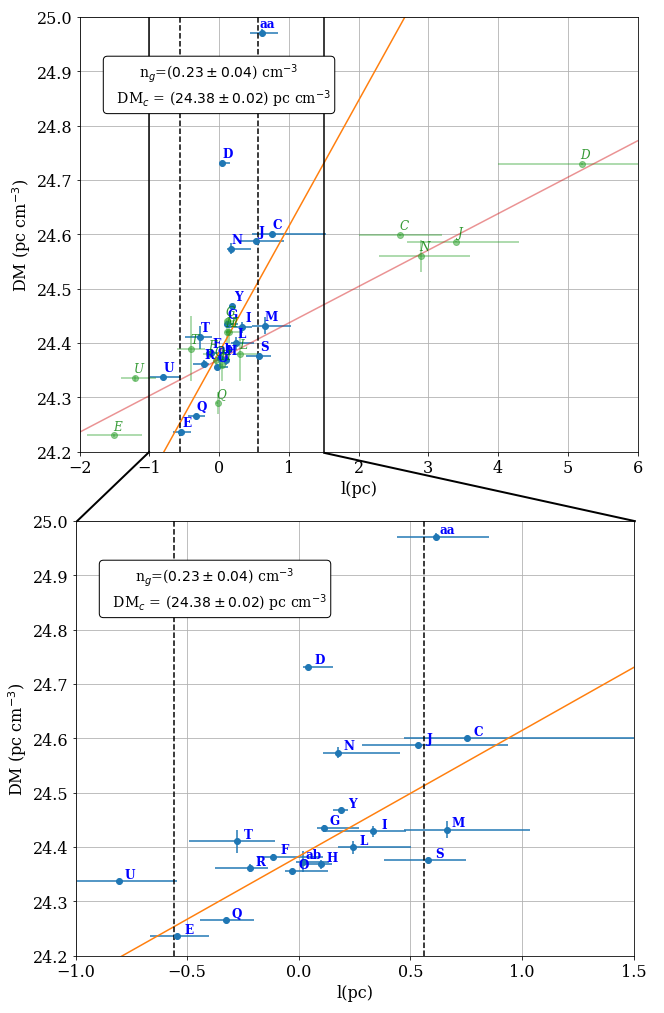}
\caption{Fit of the density of the gas assuming a model of constant density throughout the central regions of the cluster. In the top panel the blue points are associated to the position of the pulsars ($l$ being the distance of each object from the plane passing through the centre and perpendicular to the line of sight) and the error bars indicate the 68\% confidence interval on the posterior distribution for the position of the pulsars. The vertical dashed lines correspond to the core radius. The best fit is the orange line. The parameters of the best fit are shown in top left corner of the plot. In the top panel the green points are the distances along the line of sight measured by \protect\cite{Freire2001b}. The red line is the best fit found by these authors. The bottom panel is a zoom in the central region of the cluster showing only the new measurements.
}\label{constant_density_fit}
\end{center}
\end{figure}

Fig. \ref{constant_density_fit} shows the best fit with a value of the density  $n_g=0.23 \pm 0.05\, \rm{cm}^{-3}$ and $\rm{DM}_c=24.38\pm 0.02 \, \rm{pc}\, \rm{cm}^{-3}$. 
In comparison the values found by \cite{Freire2001b} are $n_g=0.067 \pm 0.015 \, \rm{cm}^{-3}$ and $\rm{DM_c}=24.381 \pm 0.009\, \rm{pc}\, \rm{cm}^{-3}$. We find a gas density which is higher than the previous estimate, although with larger uncertainty. The values of the line-of-sight positions of the pulsars and the fits for them are shown in the top panel of Fig. \ref{constant_density_fit}. The previous values are larger than those found with our analysis. This is probably because of the method used to find the line-of-sight positions of the pulsars. In particular, \cite{Freire2001b} did not fit for the cluster parameters and took an average value for the intrinsic spin-down. Moreover, since the measurements of the second derivative of the spin period were not available at the time, they had to resolve the ambiguity of the line-of-sight position arbitrarily. 

As can be seen in Fig. \ref{constant_density_fit} there are some pulsars with DM values very different from the predicted model. We tested whether these outliers had an influence on the fit. We performed the same fitting procedure using randomly chosen subsets of pulsars. In all cases the fits returned values of density and of central DM compatible with the results presented above. So we can conclude that our results are not heavily influenced by the value of some specific millisecond pulsars. 

The observed differences of the measured $\rm{DM}$ from the constant density model have a standard deviation of $\sim 0.1$ pc cm$^{-3}$. They can be considered as arising from local over-densities and under-densities inside the cluster or could be due to inhomogeneities in the interstellar medium (ISM) along the line of sight.

A possible explanation of why pulsar 47 Tuc aa is not well fitted by our model is that it could be located further back than we estimated. The probability of the pulsar being located further than 1.5 pc from the cluster centre is about 4 per cent as measured from the posterior distribution. Statistically, since we have a sample of 22 pulsars, there is a good probability that at least one pulsar would be an outlier at that level.

The region that we are able to probe with the pulsars extends to about 1 pc from the centre. Assuming that this model is valid only in this region we calculate a total mass of gas in the inner 1 pc of $0.023 \pm 0.005$ M$_{\odot}$.

\subsection{King profile distribution}

Another model to be explored is a gas density profile that follows the same King profile as the stars in the globular cluster, suitably scaled. This option is motivated by the hypothesis that  the observed ionized gas is released by the winds of massive stars. As shown in eq. (\ref{dens_3d}) with $\alpha=-2$, the gas density is:

\begin{equation}
n_g(R_{\perp},l) = n_{g,c}  \left[ 1 + \left(\frac{R_{\perp}}{r_c} \right)^2 + \left( \frac{l}{r_c}\right)^2 \right]^{-3/2},
\end{equation}
where $n_{g,c} $ is the density of the gas at the centre of the cluster.

Correspondingly, the total $\rm{DM}$ for each pulsar should be modeled by the following equation:
{}
\begin{equation}
{\rm DM}= \frac{n_{g,c} \,r_c^3}{r_c^2+R_{\perp}^2}\left( \frac{l}{\sqrt{l^2+R_{\perp}^2+ r_c^2}} + \frac{l_T}{\sqrt{l_T^2+R_{\perp}^2+ r_c^2}}\right) + \rm{DM}_f,
\end{equation}
where $\rm{DM}_f$ is the foreground contribution to the total $\rm{DM}$. 

\begin{figure}
\begin{center}
\includegraphics[width=\columnwidth]{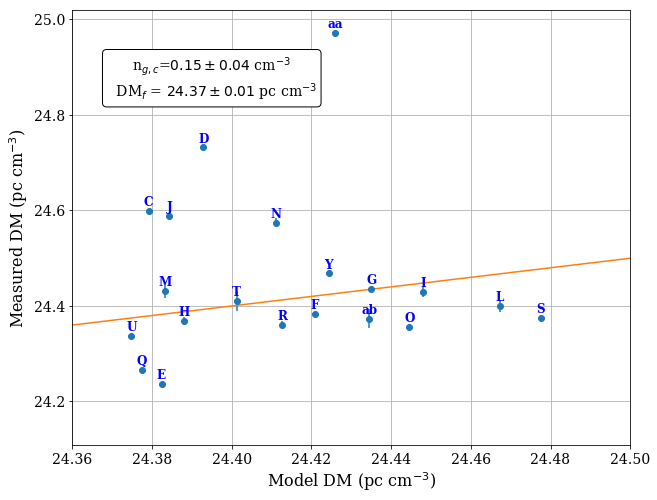}
\caption{Fit of the gas density assuming the gas is distributed following the same King profile as the stars. 
The plot shows the measured DM versus the DM predicted by the best fit. The orange line represents the unity line. In the case of a perfect fit all the points should fall on the orange line.
}\label{king_density_fit}
\end{center}
\end{figure}

The best fit for this model is shown in Fig. \ref{king_density_fit}. That implies $n_{g,c}= 0.15 \pm0.04$ cm$^{-3}$ and $\rm{DM}_f= 24.37 \pm 0.01$ pc cm$^{-3}$. 

This best fit appears to be much worse than the one reported in the previous section. The logarithm of the Bayes factor $K$ of this model with respect to the constant density model is log $K \sim 14000 >> 2$. 
This means that the model with constant gas density is strongly favoured to explain the observed data with respect to a King distribution model for the gas.

\subsection{Decreasing model}

It has also been suggested that, in the presence of an IMBH in the centre of the cluster, the gas density profile should be constantly decreasing \citep{Pepe2016}. We first tested a model in which the gas density drops as $1/r$:

\begin{equation}
n_g(R_{\perp},l) =\frac{n_{g,1}  r_c}{\sqrt{R_{\perp}^2 +l^2}},
\end{equation}
where in this case $n_{g,1}$ corresponds to the density of the gas at one core radius from the centre.

The corresponding total $\rm{DM}$ for each pulsar is:

\begin{equation}
{\rm DM} = n_{g,1}  r_c \,\rm{log}\left((l^2+ R_{\perp}^2)+l\right) + \rm{log}\left((l_T^2+ R_{\perp}^2)+l_T\right) + {\rm DM_f},
\end{equation}
where $l_T$ and $\rm{DM}_f$ are same as in the case of  a King profile.

\begin{figure}
\begin{center}
\includegraphics[width=\columnwidth]{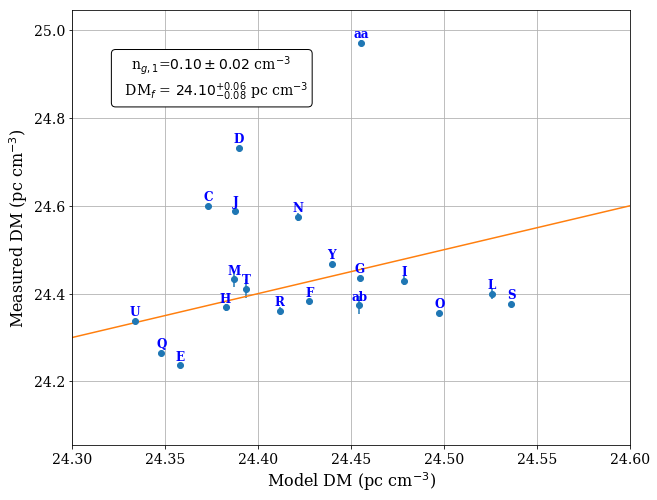}
\caption{Fit of the gas density assuming the distribution of the gas is decreasing as $r^{-1}$. 
The plot shows the measured DM versus the DM predicted by the best fit. The orange line represents the unity line. In the case of a perfect fit all the points should fall on the orange line.}
\label{decreasing_density_fit}
\end{center}
\end{figure}

The best fit for this model is reported in Fig. \ref{decreasing_density_fit} with the parameter $n_{g,1}=0.10 \pm 0.03$ cm$^{-3}$  and $\rm{DM}_f= 24.10 ^{+ 0.06}_{-0.09}$ pc cm$^{-3}$. 

In this case, the Bayes factor with respect to the constant gas density model and is log $K \sim 10000 >>2$. This means that this model with decreasing density is strongly disfavoured to explain the observed data. We repeated the exercise for a gas density scaling as $r^{-h}$ (for $h=2,3$) and always found very large values for the Bayes factor. In summary, it appears that any models in which the gas density decreases within a volume of about 1 pc from the globular cluster centre is  significantly disfavoured in comparison with a constant gas density model.
 
\section{Discussion}\label{section_discussion}

Our analysis confirms the presence of ionized gas in the central regions of 47 Tuc, as it was first reported by \cite{Freire2001b}. In addition, we have also been able to compare different distributions of gas and one with constant density in the central region is strongly favoured.

\cite{McDonald2015a} suggested that the gas might be originating from the winds of evolved RGB and AGB stars in the cluster and it could be completely ionized by the UV radiation of young white dwarfs (other kinds of stars are not effective, since they cannot provide enough ionizing photons). According to them, all the ionizing sources in the cluster support a time averaged ionizing flux of $2.43 \times 10^{47}$ photons s$^{-1}$ with a characteristic temperature of 65000 K. This radiation is enough to heat and ionize all the gas in the cluster. We can check what distribution the gas would follow in this conditions by assuming an equilibrium between the pressure forces and the gravity of the cluster.

The gravitational force on a volume element of gas is expressed by the formula:

\begin{equation}
f_g(r)= -\frac{G \rho_g(r) M_{\rm King}(r)}{r^2},
\end{equation}
where $\rho_g(r)$ is the density of the gas at radius r and $M_{\rm King}(r)$ is the total mass contained in the cluster assuming a King distribution. 

This force must be balanced by the pressure forces. There are two types of pressure: a thermal pressure caused by the temperature of the gas and a radiative pressure driven by the radiation field. This radiation field can interact with the gas since we assume that it is ionized. For an ideal gas the thermal pressure is:

\begin{equation}
P_T(r)=n_g(r) k_B T(r),
\end{equation}
where $k_B$ is the Boltzmann constant. The radiative pressure can be measured from the assumption that the radiation is a blackbody. In this case the pressure at the surface of a star is:

\begin{equation}
P_{\star}=\frac{4\sigma_{sb}}{3 c} T_R^4,
\end{equation}
where $T_R$ is the temperature of the star and $\sigma_{sb}$ is the Stefan-Boltzmann constant. The radiation pressure has a dependence on the distance $r$ from the surface as $P_R= P_{\star} (R_{\star}/r)^{2}$, where $R_{\star}$ is the radius of the star. If we assume that the radiation is coming from white dwarfs, the total radiation pressure at a radius $r$ becomes:

\begin{equation}
P_R(r)= P_{\star} \left( \frac{R_{\star}}{r} \right)^{2} N_{\star}(r),
\end{equation}
where $N_{\star}(r)$ is the number of white dwarfs contained within a radius $r$ and assuming they follow the same King distribution as the normal stars.

Therefore the pressure force per unit volume can be written as:

\begin{equation} \label{eq_din}
f_P= -\frac{dP}{dr} = -\frac{dP_T}{dr} -\frac{dP_R}{dr} 
\end{equation}

\begin{figure}
\begin{center}
\includegraphics[width=\columnwidth]{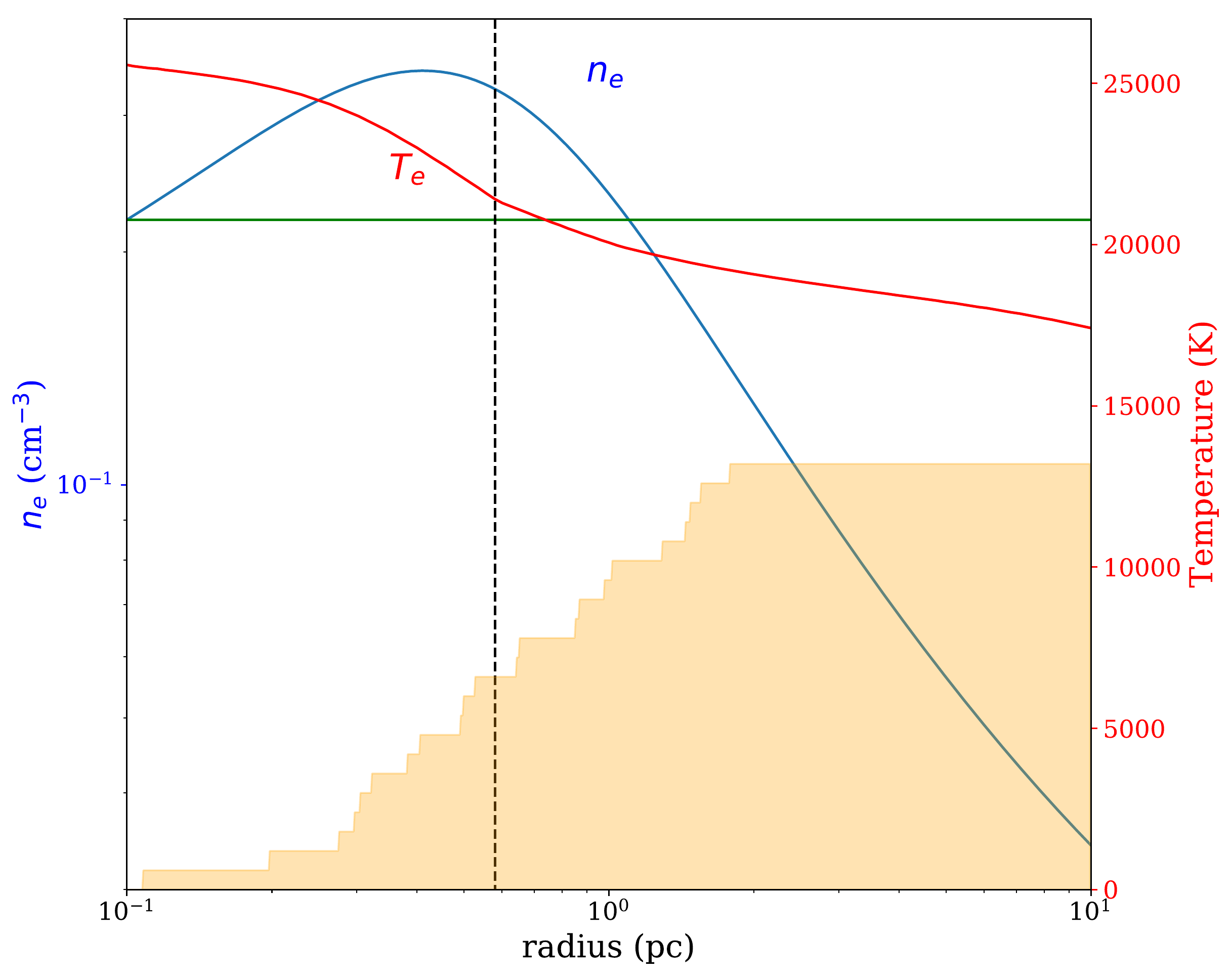}
\caption{Temperature and density profile for the gas in 47 Tuc as obtained from the {\sc cloudy} run and solving the eq. \ref{eq_din}. The dashed vertical black line is the core radius while the green horizontal line is the estimated value of $n=0.23$ cm$^{-3}$. 
The orange histogram shows the cumulative distribution of the pulsars analysed in this work. See text for more details.
}\label{Temp_dens}
\end{center}
\end{figure}

All that is needed to solve the differential equation is the temperature of the gas. The latter can be measured by solving the radiative transfer in the cluster with the given radiation field. We did this using the software {\sc cloudy}\footnote{Version 17.00 of the code is described by \cite{Ferland2017}. Software can be found at www.nublado.org.}. For the first run of the code we assumed gas at the constant density of 0.23 cm$^{-3}$ and a metallicity of [Fe/H]$=-0.72$. We used the temperature distribution found this way to solve the equilibrium equation and found the density distribution of the gas. We reiterated the process until convergence. The final temperature and density distributions are shown in Fig. \ref{Temp_dens}. The gas density appears to slightly increase in the central parsec and then drop outside. However, this distribution can hardly be distinguished from a constant density profile when looking at pulsar data, owing to the uncertainties on the line-of-sight positions and to the internal scatter of $\rm{DM}$. Moreover the set of pulsars is concentrated in the central region where the gas density has not yet decreased. The resulting temperature is able to maintain all the hydrogen and helium completely ionized and keep the heavier elements at a high ionization state. 

Many assumptions could affect the results above: e.g. the hypotheses of an ionizing radiation which is constant in time and that is produced only at the centre of the cluster could break down. A more detailed modelling of the equilibrium of the gas, including secular variations in the energy input, must be considered to better understand the behaviour and/or the status of equilibrium of the gas.

\subsection{IMBH}

\cite{Kiziltan2017a} suggested that a IMBH of mass $\sim 2200$ M$_{\odot}$ is present in the centre of 47 Tuc. 
However \cite{Freire2017} noted that the analysis that led to such claim was performed using a wrong value of the distance of the globular cluster. Using the correct value for the distance \cite{Freire2017} claim that the evidence disappears.

Interestingly, we find an increase in probability close to that same mass (see Fig. \ref{IMBH_posterior}). The increase in probability at this mass is only 4 per cent if measured by assuming a linearly decreasing background. However, since the peak of probability is at a value close to that measured by \cite{Kiziltan2017a}, we might be sensitive to the same effect.

A black hole of $\sim 2000$ M$_{\odot}$ in a cluster with a central velocity dispersion of $\sim 13$ km s$^{-1}$ would have an influence radius of only $0.05$ pc,  which is smaller than the radius of the closest pulsars. The effects of a central black hole on the accelerations could however still be visible outside the influence radius. As was shown in Fig. \ref{degeneracy} the acceleration in a King profile has a maximum value that cannot be surpassed in absence of a black hole. If a black hole is present this limit would not be so stringent and pulsars could exceed it even outside the influence radius.
As shown in Figure \ref{acc_jerk_plots} the accelerations of the pulsars are contained inside this limit taking into account the credible intervals. But a number of them have values close to this limit. Such values are more likely to be obtained in the presence of a central black hole of a certain mass. So while we found no significant evidence in favour of the presence of such black hole, the increase of probability at that mass could be explained by this phenomenon.

The discrepancy between the probability distribution of the mass of the black hole obtained in this work and in \cite{Kiziltan2017a} despite being sensitive to the same effects could be caused by different effects:
it could be due to different distances of the cluster but it could also be caused by different priors on the mass. In this work we used a logarithmic prior on the mass favouring lower values. If \cite{Kiziltan2017a} used different assumptions that favoured higher black hole masses, the peak at $\sim 2000$ M$_{\odot}$ could have gained significance.

The presence of a central black hole could further be examined via the dynamical effects on the pulsars by investigating the effects on the jerks. However predictions of how strong this effect might be are still lacking.

Our results about the shape of the gas density in the central region of 47 Tuc (a radially decreasing density profile being disfavoured with respect to a constant density distribution) do not support the presence of an IMBH at the centre but no mass limit can be given using this argument as models are not detailed enough. On the other hand, the estimate of the density and the temperature profile of the gas (see previous section) opens also the possibility of putting another independent limit on the mass of the central IMBH. If gas is present around a black hole it should accrete and emit radiation from radio to X-rays. Since only upper limits on such radiation have been derived for 47 Tuc \citep{Lu2011}, from the theory of accretion we can derive an upper limit on the mass. 

This method has been extensively used in the literature \citep{Ho2003, Maccarone2008, Lu2011} and makes use of the Bondi-Hoyle-Lyttleton theory of spherical accretion \citep{Hoyle1941, Bondi1944, Bondi1952}. According to this model, the mass accretion rate on a black hole of mass M$_{\rm IMBH}$ is:

\begin{equation}
\dot M_{\rm BHL} = 4\pi G^2 {\rm M}_{\rm IMBH}^2 \rho_g c_s^{-3},
\end{equation}
where $\rho_g$ is the density of the gas far from the black hole and $c_s$ is the sound speed far from the black hole. The sound speed for a thermal gas can be written as $c_s=\sqrt{\gamma k_B T /(\mu_{\rm mol} m_p)}$, where $\gamma=5/3$ is the adiabatic index and $\mu_{\rm mol} \sim1.25$ is the mean molecular weight. Rewriting the accretion rate as a function of the mass of the black hole and the density and temperature of the gas we obtain (the same formula in \cite{Maccarone2004} and in \cite{Lu2011} is reported with a mistake in the sign of the exponent of the temperature):

\begin{equation}
\dot M_{\rm BHL} =3.2 \times 10^{17} \left( \frac{\rm M_{\rm BH}}{2000 M_{\odot}}\right)^2 \left( \frac{n}{0.2 \, \rm{H cm}^{-3}}\right) \left( \frac{T}{10^4 \rm{K}}\right)^{-1.5} ({\rm g\, s^{-1}}).
\end{equation}

The correct value for the accretion rate $\dot m$ must account for the accretion efficiency ($\epsilon$) which is around 3\% \citep{Maccarone2008}, but can be as low as 0.1\% \citep{Ho2003}: that is because the black hole is supposed to be in a low accretion regime. This regime is prevalent in the cases where the sound speed is higher than the velocity dispersion. In our case the sound speed as measured with the formula reported above is $\sim 16$ km/s while the velocity dispersion is only $\sim 13$ km/s.

The X-ray luminosity of the black hole can be measured with the formula $ L_{\rm X} =\eta \dot m$, where $\eta$ is the radiative efficiency, which, for the low luminosity state, can be expressed as $0.5 \dot m c^4 / L_{\rm EDD}$ \citep{Maccarone2008}. In this equation ${\rm L_{EDD}}=1.26 \times 10^{38} {\rm (M_{\rm BH}/ M_{\odot})}$ erg s$^{-1}$ is the Eddington luminosity. 

This luminosity can be compared with the observation to give the maximum possible mass that avoids detection. Usually, however, the most stringent results are obtained from observations in the radio band. The flux density in radio at 5 GHz is in turn linked to the luminosity in X-rays by the following formula \citep{Merloni2003}:

\begin{equation}
 F_{\rm 5GHz} = 10 \left( \frac{L_X}{3\times 10^{31} {\rm erg\, s^{-1}}}\right)^{0.6}  \left( \frac{\rm M_{BH}}{100\, \rm{M_{\odot}}}\right)^{0.78}  \left( \frac{d}{10 \,{\rm kpc}}\right)^{-2} ({\rm \mu Jy}).
\end{equation}

Solving for ${\rm M_{BH}}$ and expressing all quantities as a function of known parameters we obtain:

\begin{equation}
\begin{split}
{\rm M_{BH}} = 44.7 \, ({\rm F_{5GHz}})^{0.39}& \left( \frac{n}{0.2 \, \rm{H cm}^{-3}}\right)^{-0.47} \left( \frac{T}{10^4 \rm{K}}\right)^{0.7} \\
 &\times\,  \epsilon^{-0.47} \left( \frac{d}{10 {\rm kpc}}\right)^{0.78} ({\rm M_{\odot}}).
\end{split}
\end{equation}

With the values for the density and temperature estimated above, the assumed distance of 47 Tuc and the 3$\sigma$ upper limit for any radio flux at 5 GHz (11.1 $\mu$Jy, \citealt{2018ApJ...862...16T}), we obtain $\sim 550 \,{\rm M}_{\odot}$ for the reference $\epsilon=3\%$ case, but also $\sim 2500\, {\rm M}_{\odot}$ for the more conservative $\epsilon=0.1\%$ hypothesis. 
In summary, the limits measured through the thermodynamic properties of the gas in 47 Tuc in the most conservative case are very close but still compatible with the claim of an IMBH of mass $\sim 2200$ M$_{\odot}$ made by \cite{Kiziltan2017a}.

\section{Conclusions}\label{section_conclusion}

In this paper we used the new timing results of the millisecond pulsars associated with the globular cluster 47 Tuc to perform a detailed modelling of the dynamics of the cluster. We measured the properties of the cluster, found an upper limit on the mass of a possible IMBH at the centre and the position along the line of sight of the pulsars. By using this information and the observed DMs of the pulsars, we tested the presence of ionized gas following different distributions. The model with the highest statistical likelihood has a constant density distribution in the region populated by the pulsars, with a density of $n_g= 0.23 \pm 0.05$ cm$^{-3}$. Other models invoking a gas density distribution that follows the stellar distribution or a radially decreasing distribution are disfavoured. 

The proposed explanation for how a region of constant gas density can be maintained in the centre of the cluster is that the thermal and the radiative pressure provide the necessary support against the gravitational collapse. However, more detailed modelling of the gas injection and of the energy input must be developed to test this model.

Finally, we used the derived information about the density and temperature profiles for the gas in order to put upper limits on the mass of  a putative IMBH at the centre of 47 Tuc. The presence of a massive central black hole in 47 Tuc will also be better constrained in the future when the effects of an IMBH on the jerks of the pulsars close enough to the latter are included.

\section*{Acknowledgements} 
We would like to thank Brian Prager and Scott Ransom for giving us access to the MCMC code and to the data on which we tested it. 
We would like to thank Monica Colpi for the useful comments and fruitful discussions. 
FA acknowledges the support of the PHAROS COST Action (CA16124).
With the support of the Italian Ministry of Foreign Affairs and International Cooperation, Directorate General for the Country Promotion (Bilateral Grant Agreement ZA14GR02 - Mapping the Universe on the Pathway to SKA).
We thank the anonymous referee for the helpful comments. 

\bibliographystyle{mnras}
\bibliography{biblio_MCMC}

\bsp	
\label{lastpage}
\end{document}